\documentclass[hyperlink,superscriptaddress,preprintnumbers,amsmath,amssymb,nofootinbib]{revtex4}
\usepackage{graphicx}
\usepackage{epstopdf}
\usepackage{dcolumn}
\usepackage{bm}
\usepackage{hyperref}

\def\slashchar#1{\setbox0=\hbox{$#1$} 
\dimen0=\wd0 
\setbox1=\hbox{/} \dimen1=\wd1 
\ifdim\dimen0>\dimen1 
\rlap{\hbox to \dimen0{\hfil/\hfil}} 
#1 
\else 
\rlap{\hbox to \dimen1{\hfil$#1$\hfil}} 
/ 
\fi}

\def\b{\beta}
\def\c{\varepsilon}
\def\d{\delta}

\def\g{\gamma}

\def\l{\lambda}
\def\m{\mu}

\def\s{\sigma}

\def\G{\Gamma}

\def\L{\Lambda}

\def\beq{\begin{eqnarray}}
\def\eeq{\end{eqnarray}}

\newcommand{\lsim}{ \mathop{}_{\textstyle \sim}^{\textstyle <} }
\newcommand{\vev}[1]{ \left\langle {#1} \right\rangle }

\def\tr{\mathop{\rm tr}}

\begin{document}
\title{Nambu--Goldstone Dark Matter and Cosmic Ray Electron and Positron  Excess } 

\author{Masahiro Ibe}
\affiliation{%
SLAC National Accelerator Laboratory, Menlo Park, CA 94025
}%
\author{Yu Nakayama}
\affiliation{%
Department of Physics, University of California, Berkeley, 
and Theoretical Physics Group, LBNL, Berkeley, CA 94720 
}%
\author{Hitoshi Murayama}
\affiliation{%
Department of Physics, University of California, Berkeley, 
and Theoretical Physics Group, LBNL, Berkeley, CA 94720 
}%
\affiliation{Institute for the Physics and Mathematics of the Universe, University of Tokyo, 
Kashiwa 277-8568, Japan.}
\author{Tsutomu T.~Yanagida}
\affiliation{%
Department of Physics, University of Tokyo, Tokyo 113-0033, Japan.}
\affiliation{Institute for the Physics and Mathematics of the Universe, University of Tokyo, 
Kashiwa 277-8568, Japan.}

\begin{abstract}
  We propose a model of dark matter identified with a
  pseudo-Nambu--Goldstone boson in the dynamical supersymmetry breaking
  sector in a gauge mediation scenario.  
  The dark matter particles annihilate via a below-threshold
  narrow resonance into a pair of R-axions each of which subsequently
  decays into a pair of light leptons. The Breit--Wigner enhancement
  explains the excess electron and positron fluxes reported in the
  recent cosmic ray experiments PAMELA, ATIC and PPB-BETS without
  postulating an overdensity in halo, and the limit on 
  anti-proton flux from PAMELA is naturally evaded.
\end{abstract}

\date{\today}
\maketitle
\preprint{IPMU09-0024}
\preprint{SLAC-PUB-13546}
\preprint{UCB-PTH-09/09}
\section{Introduction}
Existence of non-baryonic dark matter as the dominant component of
matter in the universe has been established by numerous observations.
The origin of the dark matter, however, has not been identified yet,
and its nature is arguably the most important problem in particle
and astrophysics.

Recent observations of ATIC\,\cite{Chang:2008zz} and
PPB-BETS\,\cite{Torii:2008xu} balloon experiments show the existence
of a bump in a 300-800\,GeV energy region of $e^-+e^+$ flux in cosmic
ray.  The interesting astrophysical possibilities for the origin of
the excesses are nearby
pulsars\,\cite{Hooper:2008kg,Yuksel:2008rf,Profumo:2008ms} or
supernovae remnants\,\cite{Shaviv:2009bu}.  The most exciting
interpretation of the excess, however, is the annihilation and/or
decay of the dark matter with a mass in a TeV range.  It is remarkable
that it explains simultaneously the anomalous excess of $e^+$ flux in
PAMELA experiments\,\cite{Adriani:2008zr}.

From the theoretical point of view, it is very interesting to explain
the dark matter in the supersymmetric (SUSY) extension of the standard
model (SSM), since the SUSY models naturally possess two types of
candidates for the dark matter.  One that has been discussed
extensively in the literature is the lightest SUSY particle called as
the LSP\,\cite{Murayama:2007ek}, and the other often overlooked is
stable composite ``baryons'' in a dynamical supersymmetry breaking
sector\,\cite{Dimopoulos:1996gy}.  The former case, however, implies
that the masses of the gluino and squarks are much larger than a TeV
range, which causes serious problems in discoveries of SUSY particles
at the LHC.  The latter case predicts most likely the mass of the dark
matter to be in at least about 30\,TeV, and hence, it seems difficult
to explain the dark matter with a TeV mass.

In this paper, we propose a model where the dark matter with a TeV mass
is nothing but pseudo-Nambu--Goldstone bosons generated by the SUSY
breaking sector (as in the latter case above).  Surprisingly, this model
provides solutions to the two puzzles in the recent cosmic ray
experiments.  The first puzzle is that the required annihilation cross
section in the galactic halo is much larger (by a factor of $O(100)$)
than the one appropriate to explain the dark matter relic density
precisely measured by the WMAP experiment\,\cite{Komatsu:2008hk}.  The
second puzzle is that the PAMELA experiment sees no excess in the
anti-proton flux while it sees an excess of the anti-electron flux.

As we will see, in the Nambu--Goldstone dark matter scenario, the
observed dark matter abundance is achieved only if the annihilation
process occurs near the pole of a narrow resonance.  This inevitably
evokes the Breit--Wigner enhancement of the dark matter
annihilation\,\cite{Ibe:2008ye} (see
Refs.\,\cite{Pospelov:2008jd,Feldman:2008xs} for earlier attempts)
which explains the so-called boost factor.  Furthermore, we will see
that the dominant final state of the near-pole annihilation is a pair
of the R-axions each of which subsequently decays into a light lepton
pair.  Therefore, this model also provides a concrete example of the
scenario\,\cite{Cholis:2008qq,Nomura:2008ru} explaining the second
puzzle in the PAMELA data.

The organization of the paper is as follows.  In
section\,\ref{sec:general}, we will discuss generic features of the
dark matter in the SUSY breaking sector in the light of the recent
cosmic ray experiments.  In section\,\ref{sec:model}, we propose the
Nambu--Goldstone dark matter where the dark matter annihilates via a
narrow resonance into an R-axion pair.  The final section is devoted
to our conclusion.

\section{General discussion on the Hidden Sector Dark
  Matter}\label{sec:general} 

The idea of the dark matter in the SUSY breaking sector in gauge mediation
models was first
sketched in Ref.\,\cite{Dimopoulos:1996gy}.  When the SUSY breaking
sector possesses a global symmetry, the lightest particle which is
charged under the global symmetry is stable and can be a candidate of
the dark matter.  In earlier attempts, the mass of the dark matter was
postulated to be of the order of the dynamical SUSY breaking scale,
{\it i.e.}\/, around 30\,TeV, which is the lowest possible scale realized
in gauge mediation
scenarios\,\cite{Dine:1981za,Dine:1993yw,Dine:1994vc,Dine:1995ag}.
However, such a heavy dark matter is not favorable to explain the
observed bump in a 300-800\,GeV energy region of $e^-+e^+$ flux in
cosmic ray\,\cite{Chang:2008zz,Torii:2008xu}.  Therefore, in order to
obtain a viable dark matter model in the SUSY breaking sector, we need
to consider some mechanisms to realize a ``light'' dark matter
candidates in the hidden sector.

The simplest possibility to obtain such a light particle is to
introduce a small coupling so that the dark matter candidates have a
small mass, {\it i.e.}\/,
\begin{eqnarray}
 m_{\rm DM} = \c \L_{\rm SUSY},
\end{eqnarray}
where $m_{\rm DM}$ is the mass of the dark matter which is suppressed
by a small coupling $\c$ compared with the SUSY breaking scale
$\L_{\rm SUSY}$.  The dark matter with mass in a TeV range requires
\begin{eqnarray}
 \varepsilon = 10^{-(1-2)},
\end{eqnarray}
when the SUSY breaking scale is around 30\,TeV.  However, the
introduction of the small coupling naively ends up with a too small
annihilation cross section to explain the observed dark matter
abundance.  That is, the naive estimation of the annihilation cross
section of the dark matter at the freeze-out time,
\begin{eqnarray}
 \s v_{\rm rel}&\sim& \frac{\c^{4}}{16\pi}\frac{1}{m_{\rm DM}^{2}}
 \sim \frac{1}{16\pi}
 \left(\frac{m_{\rm DM}}{\L_{\rm SUSY}}\right)^{4}
 \frac{1}{m_{\rm DM}^{2}},\cr
 &\sim& 10^{-14}\,{\rm GeV}^{-2}\times\left(\frac{m_{\rm DM}}{1\,{\rm
       TeV}} \right)^{2}
 \left(\frac{30\,{\rm TeV}}{\L_{\rm SUSY}} \right)^{4},
\end{eqnarray}
is too small to explain the observed dark matter abundance which requires,
\begin{eqnarray}
\label{eq:wmap}
\sigma v_{\rm rel} \sim 10^{-9}\,{\rm GeV}^{-2}.
\end{eqnarray}
Here, we are assuming that the final state is lighter particles in the
hidden sector which eventually decay into the SSM particles.  Besides,
since we are assuming the models with gauge mediation, the coupling
between the hidden sector and the SSM sector is rather suppressed (see
also discussion in section\,\ref{sec:model}).

A more ambitious possibility is to identify the light dark matter to the
pseudo-Nambu--Goldstone bosons resulting from a spontaneous breaking of
an approximate global symmetry in the SUSY breaking sector in analogy
with the pions in QCD.  In this case, we do not need to introduce
small couplings to realize the light dark matter.  However, the naive
estimation of the annihilation cross section of the dark matter is
again suppressed, {\it i.e.}\/,
\begin{eqnarray}
 \s v_{\rm rel} \sim   \frac{1}{16\pi}
 \left(\frac{m_{\rm DM}}{\L_{\rm SUSY}}\right)^{4}
 \frac{1}{m_{\rm DM}^{2}},
 \end{eqnarray}
 where we have assumed the breaking scale of the approximate global
 symmetry to be of the order of $\L_{\rm SUSY}$.

 Theses lessons tell us that the annihilation cross section of the
 dark matter must be enhanced than the above naive expectations.  As
 an interesting possibility, such enhancement can be realized if we
 assume that the dark matter annihilates via a narrow resonance with
 mass $M \simeq 2\,m_{\rm DM}$.  This observation, in turn, suggests
 the possible enhancement of the dark matter cross section in the
 galactic halo by the Breit--Wigner enhancement
 mechanism\,\cite{Ibe:2008ye}.  In the following section, we construct
 a model of the Nambu--Goldstone dark matter where these possibilities are realized.

\section{Nambu--Goldstone Dark Matter}\label{sec:model}

In this section, we construct an explicit model of the Nambu--Goldstone
dark matter based on a dynamical SUSY breaking model.  For that
purpose, we consider a vector-like SUSY breaking model developed in
Ref.\,\cite{Izawa:1996pk}.  As we will see, this model possesses all
the necessary ingredients to realize the Nambu--Goldstone dark matter
model where the Breit--Wigner enhancement explains the effective boost
factor and the R-axion final state explains the no excess in
anti-proton flux.

\subsection{Vector-like SUSY Breaking Model}

The vector-like SUSY breaking model is based on an $SU(2)$ gauge
theory with four fundamental representation fields
$Q_{i}(i=1,\cdots,4)$ and six singlet fields $S_{ij}=-S_{ji}$
($i,j=1,\cdots,4$)\,\cite{Izawa:1996pk}.  In this model, the SUSY is
dynamically broken when the $Q$'s and $S$'s couple in the
superpotential,
\begin{eqnarray}
 W = \l_{ij} S_{ij}Q_{i}Q_{j}, \, (i<j),
\end{eqnarray}
where $\l_{ij}$ denotes coupling constants.  
The maximal global symmetry this model may have is 
$SP(4)\simeq SO(6)$ symmetry which requires $\l_{ij} = \l$.
The SUSY is broken as a
result of the tension between the $F$-term conditions of $S$'s and
$Q$'s.  That is, the $F$-term conditions of $S_{ij}$, $\partial
W/\partial S_{ij} = \l_{ij}Q_{i}Q_{j} = 0$, contradict with the
quantum modified constraint ${\rm Pf}(M_{ij}) =\L_{\rm dyn}^{2}$ where
$M_{ij}$ denote composite gauge singlets made from $Q_i Q_j$.
Especially, when the coupling constants $\l_{ij}$ are smaller than
unity, the SUSY is mainly broken by the $F$-term of a linear
combination of the singlets $S_{ij}$.

The effective theory below the dynamical scale $\L_{\rm dyn}$ is
well-described by the gauge singlets $M_{ij}$ and $S_{ij}$ with the
effective superpotential,
\begin{eqnarray}
\label{eq:eff}
 W_{\rm eff} &=& \l_{ij}  \Lambda_{\rm dyn}
 S_{ij}M_{ij} + X  \left( {\rm Pf}(M) -  \Lambda_{\rm dyn}^{2}
 \right)  ,\,(i<j),\cr 
&=& \sum_{A =0-5}  \l_{A} \Lambda_{\rm dyn}
S_{A} M_{A} + X\left(\sum_{A = 0-5} M_{A}^{2} 
-  \Lambda_{\rm dyn}^{2}\right),
\end{eqnarray}
where $X$ is a Lagrange multiplier field which enforces the quantum
modified constraint, and we have rearranged the $S_{ij}$ and $M_{ij}$
by using appropriate linear combinations in the last expression.
Here, we have assumed that the effective composite operators $M_A$
are canonically normalized (up to order one ambiguity in the
coefficient that we will neglect in the following).%
\footnote{If we use the naive dimensional
  analysis\,\cite{Luty:1997fk}, $\L_{\rm dyn}$ is replaced with
  $\L_{\rm dyn}/4\pi$ without affecting the following discussions.}
 
Now let us assume that the SUSY breaking sector possesses an $SO(5)\subset SO(6)$
global symmetry, and take $\lambda = \lambda_{0}$ and $\lambda' =
\lambda_{a=1-5}$ with $\l<\l'$.  In this case, the lightest particle
which is charged under the $SO(5)$ symmetry is stable and can be the
dark matter candidate.  Under these assumptions, the quantum modified
constraint is solved by,
\begin{eqnarray}
\label{eq:M0}
  M_{0} = \sqrt{ \Lambda_{\rm dyn}^{2}- \sum_{a=1-5} M_{a}^{2}}.
\end{eqnarray}
By plugging $M_{0}$ into the effective superpotential in
Eq.\,(\ref{eq:eff}), we obtain
\begin{eqnarray}
\label{eq:Weff}
 W_{\rm eff} \simeq \lambda \,\L_{\rm dyn}^{2} S_{0} 
 - \sum_{a=1-5} \frac{\lambda}{2}\, S_{0}M_{a}^{2} 
 + \sum_{a=1-5}\lambda'\,\L_{\rm dyn}S_{a}M_{a} + O(M_{a}^{4}).
\end{eqnarray}
Thus, in terms of the low energy effective theory, the SUSY breaking
vacuum is given by,
\begin{eqnarray}
\label{eq:vac}
 F_{S_{0}} = \lambda \L_{\rm dyn}^{2},\quad
  S_{a} = 0,\quad
 M_{a} = 0.
\end{eqnarray}

\subsection{Dark Matter without R-symmetry Breaking}

\subsubsection{R-symmetric spectrum of the light particles}

Before introducing the R-symmetry breaking, it is worth considering
the model with no R-symmetry breaking, {\it i.e.}\/, $\vev {S_{0}} =0$, which
clarifies the necessity of the R-symmetry breaking.  In the case of
the R-symmetric vacuum, the mass spectrum is given as follows.  
First, the lightest particle which is charged under the $SO(5)$ comes from 
the scalar components of $M_{a}$ 
whose masses squared are given by,
\begin{eqnarray}
\label{eq:mDM}
 m_{\pm}^2 = (\l'^{2} \pm \l^{2})  \L_{\rm dyn}^2,
\end{eqnarray}
where the minus sign corresponds to the real component of the $M_a$ scalar.
The $\l$ dependence comes from the SUSY breaking effect coming through
the $S_{0}M_{a}^{2}$ coupling in Eq.\,(\ref{eq:Weff}).  On the other
hand, the scalar part $S_{a}$ does not receive the SUSY breaking effects,
and has the same mass with the fermion components of $S_{a}$ and
$M_{a}$, {\it i.e.}\/,
\begin{eqnarray}
\label{eq:spec}
 m_{S_{a}} = m_{M_{a}}= \l' \,\L_{\rm dyn}.
\end{eqnarray}
Therefore, we find that the dark matter is given by Re$[M_{a}]$.

The masses of the $S_{0}$ components require attention.  Since the
scalar component corresponds to a classical flat direction, its mass
vanishes at the tree-level.  The one-loop Coleman--Weinberg potential
of $S_{0}$, however, gives rise to the mass of $S_0$ as (see
appendix\,\ref{sec:cw}),
\begin{eqnarray}
\label{eq:mS0}
 m_{S_{0}} \sim \frac{\l^{3}}{(4\pi)\l'} \L_{\rm dyn}.
\end{eqnarray}
In contrast, the fermion component of $S_{0}$ contains the goldstino
which acquires a very small gravitino mass by coupling to
supergravity.

By putting it all together, we find that the masses of the dark matter as
well as the other components of $S_{a}$, $M_{a}$ and $S_{0}$ are
parametrically lighter than the dynamical scale $\L_{\rm dyn}$.  For
example, we obtain a light dark matter for small couplings,
\begin{eqnarray}
m_{\rm DM}  \simeq \c\, \L_{\rm SUSY}, \quad \c =O(\l^{1/2},\, \l'^{1/2}),
\end{eqnarray}
where we have used $\L_{\rm SUSY} = \l^{1/2} \L_{\rm dyn}$ and $\l \lesssim \l'$.
Thus, the dark matter with a mass in a TeV range can be achieved for
\begin{eqnarray}
\label{eq:small}
 \c = 10^{-(1-2)}, \quad (\l\,,\l' = 10^{-(2-4)}).
\end{eqnarray}

The above spectrum also poses the other possibility discussed in the
previous section, {\it i.e.}\/, the pseudo-Nambu Goldstone boson dark matter.
We can see it by taking the limit of $SO(6)$ global symmetry,
{\it i.e.}\/, $\l\to \l'$.  There, the mass of the dark matter in
Eq.\,(\ref{eq:mDM}) vanishes.  This shows that the dark matter is
nothing but the pseudo-Nambu--Goldstone boson of the spontaneous
breaking of $SO(6)_{\rm app}\to SO(5)$ with a breaking scale $\L_{\rm
  dyn}$ (see Eq.\,(\ref{eq:M0})).  In this case, we obtain the dark
matter with a TeV mass for
\begin{eqnarray}
\label{eq:degenerate}
 \l'-\l = O(10^{-(2-4)}),
\end{eqnarray}
while keeping $\l$ and $\l'$ of the order of one.


\subsubsection{Dark matter annihilation without resonance}
Now, let us consider the dark matter annihilation.  For,
$m_{S_{0}}<m_{\rm DM}$, the dark matter Re$[M_{a}]$ dominantly
annihilates into $S_{0}$ scalar via the $F$-term potential $|m_{S_{a}}
S_{a}-\l S_{0}M_{a}|^2$ (see Eq.\,(\ref{eq:Weff})).  The amplitude of
this process is given by,
\begin{eqnarray}
 {\cal M} = \l^{2} + \l^{2}\frac{ m_{S_{a}}^{2}}{t-m_{S_{a}}^{2} } =
 \l^{2}\frac{t}{t-m_{S_{a}}^{2}}, 
\end{eqnarray}
where $t $ denotes the momentum transfer.  The first term comes from
the four-point interaction and the second term from the $t$-channel
exchange of the $S_{a}$ scalars.  In the $S$-wave limit, the momentum
transfer is given by,
\begin{eqnarray}
 t  = - m_{\rm DM}^{2} \beta_{f}^{2}, \quad \beta_{f} = \sqrt{1-
   \frac{m_{S_{0}}^{2} }{m_{\rm DM}^{2}   }}  \simeq 1,
\end{eqnarray}
and the cross section is given by,
\begin{eqnarray}
\label{eq:cs1}
\sigma v_{\rm rel} &=& \frac{\b_f}{8\pi}\frac{v_{\rm rel}}{2 (2 m_{\rm
    DM})^{2} \beta_{i}}  \l^{4} \left(\frac{t}{t-m_{S_{a}}^{2}}\right)^{2},\cr
&\simeq &\frac{\l^{4}}{32\pi} \frac{\b_{f}^{5}}{m_{\rm DM}^{2} }
\left( \frac{m_{\rm DM}^{2}}{m_{S_{a}}^{2}} 
 \right)^{2}.
\end{eqnarray}
where the final approximation is valid for $m_{\rm DM}\lsim m_{S_{a}}$.%
\footnote{For $m_{S_{0}}>m_{\rm DM}$, the dark matter dominantly annihilates
into the gravitinos with a much more suppressed annihilation cross section.}

From this expression, we confirm that the cross section of the
``light'' dark matter, i.e. 
 $\l,\l' = 10^{-(2-4)}$ (Eq.\,(\ref{eq:small}))
or $m_{\rm DM}\simeq 1$\,TeV and $m_{S_a}\simeq 30$\,TeV
(Eq.\,(\ref{eq:degenerate})),
is highly suppressed.
So, we need to look for
an appropriate narrow resonance so that the cross section is
sufficiently enhanced.  Interestingly, in the case of the Nambu--Goldstone dark
matter, there is a candidate for such a resonance, the scalar part of
$S_{0}$.  The Eq.\,(\ref{eq:mS0}) shows that the mass of $S_{0}$
scalar can be also in a TeV range for $\l \lesssim 1$, and hence, the
$S_{0}$ mass can satisfy $m_{S_{0}}\simeq 2\,m_{\rm DM}$ with a
careful tuning.  Thus, if the dark matter annihilates via the $S_{0}$
resonance, the annihilation cross section can be drastically enhanced from the
one given above.  However, for this process, the R-symmetry must be
broken, since the R-charge of $S_{0}$ is $2$, while that of $M_{a}$
is 0.  Motivated by these observations, we will extend our analysis
to the model with the R-symmetry breaking.

\subsection{Nambu--Goldstone Dark Matter with R-symmetry breaking}

\subsubsection{R-symmetry breaking}

Now, let us consider spontaneous R-symmetry breaking.
For simplicity, we assume that the R-symmetry is broken by effects of 
higher dimensional operators of $S_{0}$ in the K\"ahler potential, 
\begin{eqnarray}
\label{eq:Kahler}
 K  &=& |S_{0}|^{2}+\frac{|S_{0}|^{4}}{4\Lambda_{4}^{2}}
 -\frac{|S_{0}|^{6}}{9\Lambda_{6}^{4}} + \cdots,
\end{eqnarray} 
where $\Lambda$'s denote the dimensionful parameters and the ellipsis
denotes the higher dimensional terms of $S_{0}$.  
The positivity of the coefficient of the quartic term is crucial to destabilize the
R-symmetric vacuum at $S_0 = 0$.
Notice that the
above K\"ahler potential provides an effective description of a quite
general class of the models with spontaneous breaking of the
R-symmetry breaking.  Especially, when the above K\"ahler potential
results from radiative corrections from physics at the scale
$\L_{\rm dyn}$, the dimensionful parameters are expected to be,
\begin{eqnarray}
\label{eq:dim4}
 \frac{1}{\L_{4}^{2}} = \frac{c_{4}^{2}}{16\pi^{2}}\frac{1}{\L_{\rm dyn}^{2}},\quad
  \frac{1}{\L_{6}^{4}} = \frac{c_{6}^{2}}{16\pi^{2}}\frac{1}{\L_{\rm dyn}^{4}},
\end{eqnarray}
where dimensionless coefficients $c_{4,6}$ are of the order of 
unity.  In appendix\,\ref{sec:mason}, we demonstrate an explicit
perturbative model which breaks the R-symmetry in a similar way
studied in Ref.\,\cite{Dine:2006xt}.

From the above K\"ahler potential, the R-symmetry is spontaneously
broken by the vacuum expectation value of the scalar component of $S_{0}$;
\begin{eqnarray}
\vev{S_{0}} &=& \frac{1}{\sqrt{2}} \frac{\Lambda_{6}^{2}}{\Lambda_{4}}
= \frac{1}{\sqrt{2}}\frac{c_{4}}{c_{6}}\L_{\rm dyn}
= \frac{1}{\sqrt{2}}f_{R},
\end{eqnarray}
where we have introduced the R-symmetry breaking scale $f_{R} =
O(\L_{\rm dyn})$ and define the R-symmetry so that $\vev {S_0}>0$.  At
this vacuum, the scalar component of $S_{0}$ is decomposed into a
flaton $s$ and the R-axion $a$ by,
\begin{eqnarray}
  S_{0} = \frac{1}{\sqrt{2}} (f_{R}+s) e^{i a/f_{R}}.
\end{eqnarray}
Then, the mass of the flaton is given by,
\begin{eqnarray}
\label{eq:mflaton}
 m_{s} = 4\sqrt{2}\, 
\frac{\l\L_{\rm dyn}^{2}\L_{4}^{3}}{(4\L_{4}^{4}+ \L_{6}^{4})}
\simeq \sqrt{2} \frac{\l \L_{\rm dyn}^{2}}{\L_{4}}
\simeq \sqrt{2} \frac{c_{4}}{4\pi} \l \L_{\rm dyn},
\end{eqnarray}
where we have used $F_{S_{0}} = \l \L_{\rm dyn}^{2}$ and assumed
Eq.\,(\ref{eq:dim4}) with $c_{4}= c_{6}=O(1)$.  Therefore, the flaton
can be in a TeV range for $\l\sim 1$ and $c_{4}\sim 1$, which is a
crucial property for the flaton to make the narrow resonance
appropriate for the dark matter annihilation.

On the other hand, the R-axion mass is much more suppressed and mainly
comes from the constant term in the superpotential which breaks the
R-symmetry explicitly.%
\footnote{In this study, we assume that the messenger sector of the
  gauge mediation also respects the R-symmetry.  Otherwise, the
  radiative correction to the K\"ahler potential of $S_0$ from the
  messenger sector gives rise to the dominant contribution to the
  R-axion mass.  
  The R-breaking mass from the Higgs sector, on the other hand, 
  is smaller than the one in Eq.\,(\ref{eq:Rmass}), 
  even if the so-called $\mu$-term does not respect the R-symmetry. 
  } 
  In the supergravity with (almost) vanishing
cosmological constant, the R-axion acquires a small
mass\,\cite{Bagger:1994hh},
\begin{eqnarray}
\label{eq:Rmass}
m_{\mathrm{axion}}^2 \sim \frac{m_{3/2} F_{S_{0}}}{f_{R}}.
\end{eqnarray}
In the case of the low-scale gauge mediation with the dynamical SUSY
breaking scale around 30\,TeV, the R-axion mass is tens to
hundreds of MeV range.

\subsubsection{Spectrum and interactions of light particles}

The spectrum of other light particles becomes also complicated in the
presence of the R-symmetry breaking, since $S_{a}$ and $M_{a}$ scalars
mix with each other via a cross term in the $F$ -term potential $|
m_{S_{a}} S_{a} -\l S_{0} M_{a} |^{2}$.  To analyze the mass spectrum
and interactions of those particles, we decompose $M_{a}$ and $S_{a}$
as
\begin{eqnarray}
  S_{a} &=& \frac{1}{\sqrt 2} (x_{s}+i\, y_{s})e^{i a/f_{R}}, \cr
  M_{a} &=& \frac{1}{\sqrt 2} (x_{m}+i\, y_{m}).
\end{eqnarray}
Here, we have suppressed the index $a$, since the particles with 
different values of $a$ decouple from each other in the following
analysis.

By using this expressions, we obtain a scalar potential,
\begin{eqnarray}
\label{eq:scalarDM}
V &=& | \l' \L_{\rm dyn} \,M_{a}|^{2} + \left| \l \L_{\rm dyn}^{2} -
  \frac{\l}{2}M_{a}^{2}\right|^{2} 
+ | \l S_{0} M_{a} + \l'\L_{\rm dyn} S_{a} | ^{2},\cr
&=& \frac{1}{2}\left( (\l'^{2}-\l^{2})\L_{\rm dyn}^{2}
  +\frac{\l^{2}}{2} (s+f_{R})^{2}\right) x_{m}^{2}  
+ \frac{1}{2}\left( (\l'^{2}+\l^{2})\L_{\rm dyn}^{2} +\frac{\l^{2}}{2}
  (s+f_{R})^{2}\right) y_{m}^{2}\cr 
&&+\,\frac{1}{2}\l'^{2} \L_{\rm dyn}^{2} x_{s}^{2}+\frac{1}{2}\l'^{2}
\L_{\rm dyn}^{2} y_{s}^{2}  
+ \frac{\l \l'}{\sqrt{2}} \L_{\rm dyn} (s+f_{R})\, x_{m} x_{s} +
\frac{\l \l'}{\sqrt{2}} \L_{\rm dyn} (s+f_{R})\, y_{m} y_{s}\cr 
&&+ \,\l^2\L_{\rm dyn}^4 + \frac{\l^2}{2^4}(x_m^2+y_m^2)^2.
\end{eqnarray}
Notice that the R-axion does not show up in the scalar interactions in
this basis, and it only appears in the derivative couplings.  From
this potential, we find that the pseudo-Nambu--Goldstone mode resides
not in $(y_{m}, y_{s})$ but in $(x_{m}, x_{s})$.  In the following, we
concentrate on the real parts $(x_{m}, x_{s})$.

The mass-squared matrix of $(x_{m},x_{s})$ is given by,
\begin{eqnarray}
 M^{2} = 
\left(
\begin{array}{cc}
 (\l'^{2}-\l^{2})\L_{\rm dyn}^{2} +\l^{2} \vev{S_{0}}^{2}   & \l \l'
 \L_{\rm dyn} \vev{S_{0}}   \\ 
  \l \l' \L_{\rm dyn} \vev{S_{0}}   &    \l'^{2}\L_{\rm dyn}^{2}
\end{array}
\right),
\end{eqnarray}
and hence,  the masses of the eigenmodes ($\phi$, $H$) are;
\begin{eqnarray}
 m_{\phi}^{2} &=&\frac{1}{2} \left(\tr M^{2}- \sqrt{ (\tr
     M^{2})^{2}-4\det M^{2}}  \right)=\frac{\det M^{2}}{m_{H}^{2}},\\ 
 m_{H}^{2} &=& \frac{1}{2} \left(\tr M^{2}+ \sqrt{ (\tr
     M^{2})^{2}-4\det M^{2}}  \right),\\ 
 \tr M^{2} &=& (2\l'^{2}-\l^{2}) \L_{\rm dyn}^{2} + \l^{2}\vev{S_{0}}^{2},\\
  \det M^{2} &=& \l'^{2}( \l'^{2}-\l^{2})\L_{\rm dyn}^{4}.
\end{eqnarray}
The mixing angle is given by,
\begin{eqnarray}
 x_{m} &=& \cos\theta\, \phi -  \sin\theta\, H,\cr
 x_{s} &=& \sin\theta\, \phi +  \cos\theta\, H,
\end{eqnarray}
with
\begin{eqnarray}
\tan\theta &=& -\frac{\l\l' \L_{\rm dyn}\vev{S_{0}}}{\l'^{2}\L_{\rm
    dyn}^{2}-m_{\phi}^{2}},\cr 
\sin\theta\cos\theta &=& -\frac{\l\l'\L_{\rm dyn} \vev{S_{0}}
}{m_{H}^{2}-m_{\phi}^{2}}. 
\end{eqnarray}
As a result, we find that the lighter scalar $\phi$ denotes the
Nambu--Goldstone mode in the limit of $\l = \l'$, and hence, we
consider $\phi$ as the dark matter.

The R-axion interactions only appear in the kinetic terms.  In the
basis we have defined, the R-axion interaction comes from the kinetic
terms of $S_{0}$ and $S_{a}$,
\begin{eqnarray}
\label{eq:kin}
{ \cal L } = \frac{1}{2} (\partial a)^{2}\left( 1 + \frac{s}{f_{R}} \right)^{2}
+ \frac{1}{2 f_{R}^{2} } (\partial a)^{2}(x_{s}^{2}+y_{s}^{2} )
+ \frac{1}{f_{R}} \partial_{\m} a (x_{s}\partial^{\m} y_{s}  - y_{s}\partial^{\m} x_{s}).
\end{eqnarray}

Altogether, in the Nambu--Goldstone dark matter scenario ({\it i.e.}\/, $\l'-\l
\ll 1$), light particles sector consists of the dark matter and the
flaton in a TeV range, and the gravitino and the R-axion with much
smaller masses, while the other components in $S_{a}$ and $M_{a}$
have masses of the order of the SUSY breaking scale.  The most
relevant terms for the dark matter annihilation is, then, given by,
\begin{eqnarray}
\label{eq:int}
{\cal L}_{\rm int} = \frac{\l^{2}}{2} f_{R} \frac{m_{\phi}^{2}} {m_{H}^{2}-m_{\phi}^{2}}\, s\, \phi^{2}
+  \frac{1}{2} (\partial a)^{2}\left( 1 + \frac{s}{f_{R}} \right)^{2},
\end{eqnarray}
where the first term comes from the scalar potential in
Eq.\,(\ref{eq:scalarDM}), while the second term comes from
Eq.\,(\ref{eq:kin}).

\subsubsection{Flaton decay}
In order to discuss the dark matter annihilation via the $s$-channel
exchange of the flaton, it is important to know the decay properties
of the flaton.  In particular, the decay rate into a dark matter pair is
important even if the pole is unphysical, {\it i.e.}\/, $2m_{\phi}>m_{s}$,
since the decay rate must be defined not on the exact pole, but on the
center of mass energy of the dark matter collision, $E_{\rm CM}$.

First, we consider the decay mode into a pair of the R-axions.  The
relevant interactions of the decay comes form the first term in
Eq.\,(\ref{eq:int}), and the decay rate into a pair of the R-axion is
given by,
\begin{eqnarray}
\label{eq:saa}
\G_{s\to aa} = \frac{1}{32\pi} \frac{m_{s}^{3}}{f_{R}^{2}},
\end{eqnarray}
where we have neglected the mass of the R-axion and taken the final
state velocity to be $\b_{f}=1$.  For example, the decay rate is very
small, {\it i.e.}\/,  $\G/m\lsim 10^{-4}$ for $f_{R}\gtrsim 30$\,TeV and $m_{s}
= 2$\,TeV.  As we will see this is favorable to realize a large
effective boost factor via the Breit--Wigner enhancement.

Next, we consider the flaton decay into a pair of the dark matter.
The relevant interaction term is given in Eq.\,(\ref{eq:int}) and the
resultant decay rate is given by,
\begin{eqnarray}
 \G_{s\to \phi\phi} = \frac{\beta_{\phi}}{32\pi}\frac{\l^{4}f_{R}^{2}}{m_{s}^{2}} \left(\frac{m_{\phi}^{2}}{m_{H}^{2}-m_{\phi}^{2}} \right)^{2} m_{s},
\end{eqnarray}
where $\beta_{\phi}$ denotes the size of the velocity of the dark
matter.  Notice that the value of $\G_{s\to \phi\phi}/\beta_{\phi}$ is
well-defined even in the unphysical region, {\it i.e.}\/, $2 m_{\phi}> m_{s}$.
The value of $\G_{s\to \phi\phi}/\beta_{\phi}$ is at most comparative
to $\G_{s\to aa}$,
\begin{eqnarray}
 \G_{s\to \phi\phi} \simeq \frac{\beta_{\phi}}{512\pi} \left(\frac{\l f_{R}}{m_{H}}\right)^{4} 
 \frac{m_{s}^{3}}{f_{R}^{2}},
\end{eqnarray}
where we have used $m_{\phi}\simeq m_{s}/2$ and $m_{H}\gg m_{\phi}$.
Therefore, we find that the decay rate into a dark matter pair does
not dominate over the one into an R-axion pair.

Let us also consider the flaton decay into a pair of the gravitinos.
The relevant interaction comes from the higher dimensional terms in
the K\"ahler potential Eq.\,(\ref{eq:Kahler}), and the resultant
interaction term is given by,
\begin{eqnarray}
{ \cal L}_{\rm int}\sim \frac{F_{S_{0}}}{\L_{4}^{2}}\, s\, \psi\psi + h.c.
 = \frac{m_{s}^{2} }{F_{S_{0}}} \, s\, \psi\psi + h.c. ,
\end{eqnarray}
where we have used Eq.\,(\ref{eq:mflaton}), {\it i.e.}\/, $\L_{4}\sim
F_{S_{0}}/m_{s}$.  Therefore, the decay width is suppressed by
$(m_{s}/\L_{\rm SUSY})^{4}$, and hence, this mode is further
suppressed compared with the mode into an R-axion pair.

Putting them all together, we obtain the flaton decay width at $E_{\rm
  CM}\simeq m_{s}$,
\begin{eqnarray}
 \G_{s} (E_{\rm CM}) = \G_{s\to aa} + \G_{s\to \phi\phi} + \cdots.
\end{eqnarray}
where $E_{\rm CM}>2 m_{\phi}$, and the dots refer to the modes into
the MSSM particles (see appendix\,\ref{sec:ssm}).  In the
following analysis, we approximate the above decay rate by,
\begin{eqnarray}
 \G_{s} (E_{\rm CM}) \simeq \G_{s}(m_{s})\simeq \G_{s\to aa},
\end{eqnarray}
since all the other modes are subdominant at $E_{\rm CM}\simeq m_{s}$.

\subsubsection{Dark matter annihilation via the $s$-channel flaton}
Now, let us consider the dark matter annihilation via the $s$-channel
flaton exchange.  The relevant interactions are again given in
Eq.\,(\ref{eq:int}).  The amplitude of this process is given by
\begin{eqnarray}
{\cal M} = \l^{2} \frac{m_{\phi}^{2}}{m_{H}^{2}-m_{\phi}^{2}}
\frac{E_{\rm CM}^{2}} {E_{\rm CM}^{2}-m_{s}^{2} + i m_{s}\G_{s}(E_{\rm
    CM})}\ ,
\end{eqnarray}
and the cross section by
\begin{eqnarray}
\label{eq:DMCS}
\s v_{\rm rel} &=& 
\frac{v_{\rm rel}}{32\pi} \frac{\b_{f}}{\b_{\phi}}
\left( \frac{m_{\phi}^{2}}{m_{H}^{2}-m_{\phi}^{2}}\right)^{2}
\frac{\l^{4}E_{\rm CM}^{2}} { (E_{\rm CM}^{2}-m_{s}^{2})^{2} + m_{s}^{2}\G_{s}^{2}}\cr
&\simeq &
\frac{\l^{4} }{64\pi} 
\left( \frac{m_{\phi}}{m_{H}}\right)^{4} \frac{1}{m_{\phi}^{2}}
\frac{1}{ (\delta + v_{\rm rel}^{2}/4)^{2} + \g_{s}^{2}}\ ,
\end{eqnarray}
where the $\G_{s}$ and $\b_{\phi}$ are defined at $E_{\rm CM}>2m_{\phi}$.
In the final expression, we have used the non-relativistic approximation,
\begin{eqnarray}
  E_{\rm CM}^{2} = 4 m_{\phi}^{2} + m_{\phi}^{2} v_{\rm rel}^{2},
\end{eqnarray}
and introduced parameters $\d$ and $\g_{s}$ by
\begin{eqnarray}
m_{s}^{2}=4m_{\phi}^{2}(1-\delta),\quad \g_{s} = \G_{s}/m_{s}.
\end{eqnarray}
From this expression, we find that the annihilation cross section of
the dark matter is substantially enhanced compared with the one given in
Eq.\,(\ref{eq:cs1}), for $|\d|, \g_{s}\ll 1$, which allows 
a sufficient annihilation cross section to reproduce the observed
dark matter density.

\subsubsection{Dark matter density and Breit--Wigner enhancement}

\begin{figure}[t]
 \begin{minipage}{.45\linewidth}
  \includegraphics[width=\linewidth]{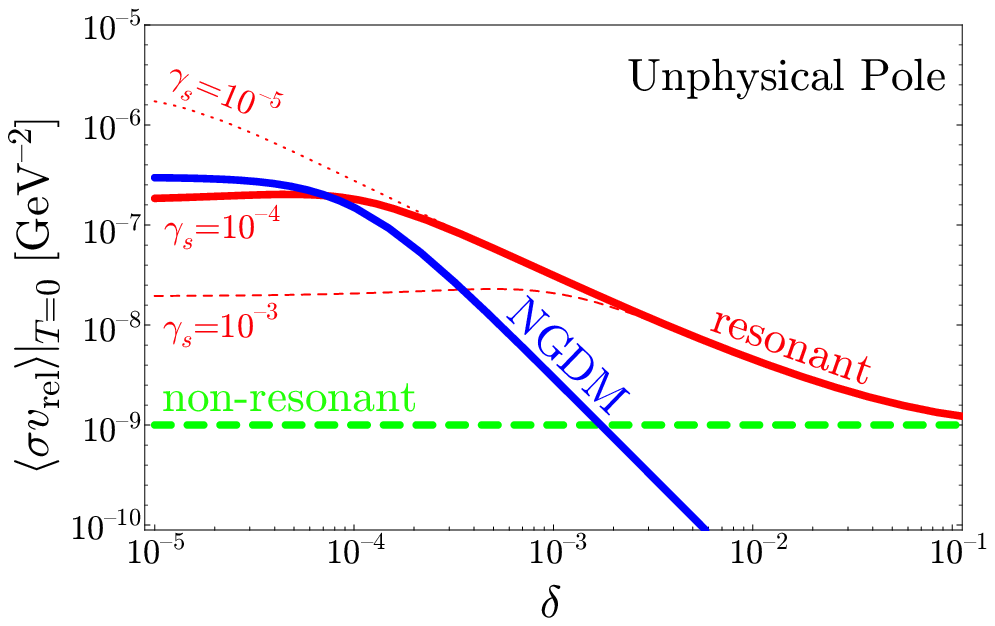}
 \end{minipage}
 \begin{minipage}{.45\linewidth}
  \includegraphics[width=1.0\linewidth]{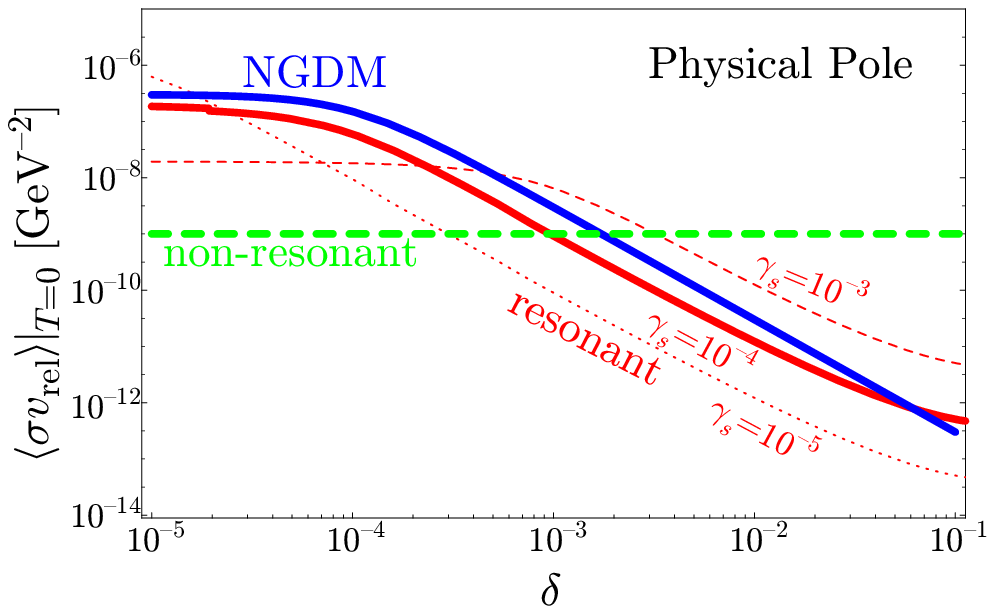}
 \end{minipage}
 \caption{Left) The $\delta$ dependence of the required annihilation
   cross section at zero temperature from the observed dark matter
   density in the case of the unphysical pole.  The red lines
   correspond to $\g_{s}=10^{-3}$, $\g_{s}=10^{-4}$ and
   $\g_{s}=10^{-5}$ from bottom to top.  The green line shows the
   required annihilation cross section in the usual thermal history.
   The blue line shows the predicted annihilation cross section for
   $\l=0.84$, $m_{H}=30\,$TeV and $m_{\phi}=1$\,TeV.  Right) The required
   annihilation cross section in the case of the physical pole.  }
\label{fig:CS}
\end{figure}
Although we obtained the enhanced annihilation cross section of the
dark matter, we should note that the thermal history of the dark
matter density is drastically changed from the usual thermal relic
density when the dark matter annihilates via the narrow
resonance\,\cite{Griest:1990kh,Gondolo:1990dk}, and hence, the
required annihilation cross section is different from the value given
in Eq.\,(\ref{eq:wmap}).  Instead, in terms of the annihilation cross
section at the zero temperature, the required annihilation cross
section to obtain the correct abundance is given by\,\cite{Ibe:2008ye},
\begin{eqnarray}
 \vev{\s v_{\rm rel}}|_{T = 0} \sim 10^{-9}\,{\rm GeV}^{-2} \times \frac{x_{b}}{x_{f}}.
\end{eqnarray}
Here $x_{f}\simeq 20$ denotes the freeze-out parameter of the usual
(non-resonant) thermal freeze-out history, while $x_{b}$ is defined
by,
\begin{eqnarray}
\frac{1}{x_{b}} \simeq \frac{1}{\vev{\s v_{\rm rel}}|_{T=0} }
\int_{x_{f}}^{\infty}\frac{\vev{\s v_{\rm rel}}}{x^{2}} \,dx.
\end{eqnarray}

In the case of the unphysical
pole, {\it i.e.}\/, $m_{s}<2 m_{\phi}$, $x_{b}$ is well approximated by
min[$\d^{-1}$,$\g_{s}^{-1}$], and the above required annihilation
cross section at the zero temperature is simply given by,
\begin{eqnarray}
\label{eq:wmap2}
\vev{ \sigma v_{\rm rel}}|_{T=0} \sim 10^{-9}\,{\rm GeV}^{-2} \times 
\frac{1}{x_{f}\,{\rm Max}[\delta,\gamma_{s}]}.
\end{eqnarray}
In Fig.\,\ref{fig:CS}, we show the required annihilation cross section
for given parameters as red lines.  The figure shows that
Eq.\,(\ref{eq:wmap2}) gives a good approximation.

On the other hand, in the case of the physical pole, the estimation of
$x_{b}$ is much more complicated.  In particular, the thermal average
picks up the pole at $v_{\rm rel}^{2}=4|\d|$ when the temperature is
rather high, {\it i.e.}\/, $x^{-1}\gg|\d|$, and hence, the annihilation cross
section can be higher at the higher temperature than the one at the
zero temperature.  As a result, the required annihilation cross
section at the zero temperature can be much lower than the one in the
usual thermal relic history.  In Fig.\,\ref{fig:CS}, we also show the
required cross section at the zero temperature in the case of the
physical pole.  The figure shows that the required cross section can
be lower than the usual value.

Now, let us compare these values with the dark matter annihilation
cross section given in Eq.\,(\ref{eq:DMCS}).  For example, if we take,
$m_{\phi}=1$\,TeV, $f_{R}=30$\,TeV, $m_{H} = 30$\,TeV and $\l = 1$,
the decay rate is very small $\g_{s}\simeq 10^{-4}$.  In this case,
the cross section at the zero temperature is
\begin{eqnarray}
\label{eq:DMCS2}
\vev{\sigma v_{\rm rel}}|_{T=0} \simeq 3\times 10^{-15}\,{\rm GeV}^{-2}\times 
\frac{\l^{4}}{\delta^{2}+\g_{s}^{2}}
\left(\frac{m_{\phi}}{1\,{\rm TeV}}\right)^{2}
\left(\frac{30\,{\rm TeV}}{m_{H}}\right)^{4}.
\end{eqnarray}
In  Fig.\,\ref{fig:CS}, we show the predicted annihilation cross
section.  From the figure, we find that the required annihilation
cross section is achieved at
\begin{eqnarray}
 \delta &\sim& 10^{-4}, \quad (\mbox{for unphysical pole}),\nonumber \\
  \delta &\sim& -10^{-1}, \quad (\mbox{for physical pole}),
\end{eqnarray}
for the given parameter set ($\l = 1$, $m_{\phi} =1$\,TeV, and
$m_{H}=f_{R}=30$\,TeV).  Therefore, the Nambu--Goldstone dark matter is
consistent with the observed dark matter density when it annihilates
via the flaton resonance with the values of $\delta$ given above.

Interestingly, the Nambu--Goldstone dark matter predicts a non-trivial
effective boost factor.  The effective boost factor in the
Breit--Wigner enhancement is defined by,
\begin{eqnarray}
{\rm BF} = \frac{x_{b}}{x_{f}}.
\end{eqnarray}
Thus, the effective boost factor for the above two solutions are given by,
\begin{eqnarray}
 {\rm BF} &\sim& 10^{2}, \quad (\mbox{for unphysical pole}),\nonumber \\
 {\rm BF} &\sim& 10^{-3}, \quad (\mbox{for physical pole}),
\end{eqnarray}
respectively for the parameters given above.

Therefore, we find that the Nambu--Goldstone dark matter model predicts
a non-trivial effective boost factor.  Especially, the model with the
unphysical flaton pole ($m_s < 2 m_\phi$) is strongly favored in the
light of the recent cosmic ray experiments.  In this case, the
parameter dependence of the boost factor is simply given by,
\begin{eqnarray}
{\rm BF} \sim 10^{2} \times  \l^4 \left(\frac{m_{\phi}}{1\,{\rm TeV}}\right)^{2}
\left(\frac{30\,{\rm TeV}}{m_{H}}\right)^{4}.
\end{eqnarray}
Here, we have used Eqs.\,(\ref{eq:DMCS}) and (\ref{eq:DMCS2}).

\subsubsection{R-axion decay and anti-proton flux}
As we have discussed, the dark matter dominantly annihilates into an
R-axion pair via the flaton resonance.  Interestingly, since the
R-axion has a mass in the range of tens to hundreds of MeV, it mainly decays
into light lepton pairs (see Ref.~\cite{Goh:2008xz} for detailed
discussion on the R-axion properties).

Therefore, the Nambu--Goldstone dark matter model provides a concrete
example of the scenario developed in Ref.\,\cite{Cholis:2008qq}.  There,
the dark matter annihilate into a new light particle which
subsequently decays into light leptons.  In this way, we can obtain a
hard positron spectrum without any additional anti-protons, so that we
can explain the PAMELA results consistently.

Here, we comment on the constraints on the decay constant and mass of
the R-axion.  For the R-axion in a mass range between two electrons
and two muons, the stringent constraint comes from a beam-dump
experiment\,\cite{Bergsma:1985qz}, which constrains the decay
constant as
\begin{eqnarray}
 f_R \gtrsim 10^{4.5}\,{\rm GeV}\times \left( \frac{m_a} {10\,\rm MeV}
 \right)^{1/2}, 
\end{eqnarray}
when we assume that the Higgs sector respects the R-symmetry and the
R-charge of the so-called $\mu$-term, $H_uH_d$, is
two\,\cite{Goh:2008xz}.  Thus, our choice of the scales of the
R-symmetry breaking and the SUSY breaking in the previous discussion
are marginally consistent with the constraint.%
\footnote{For other choices of the R-charge of the $\mu$-term, the
  constraint can be changed.  For example, when the R-charge of the
  $\mu$-term is zero, the R-axion does not mix with the neutral Higgs
  bosons in the SSM.  In this case, the couplings between the R-axion
  and the SM fermion vanish at the tree-level, and hence, the above
  constraint on the decay constant is weakened.  We may also consider
  the Higgs sector without the R-symmetry.  In that case, the degree
  of the mixing between the R-axion and the Higgs bosons is also
  altered from the one discussed in Ref.\,\cite{Goh:2008xz}, which may
  weaken the above constraint.  } On the other hand, for the R-axion
heavier than two muons, the most stringent constraint comes from the
rare decay of the $\Upsilon$ meson, $Br(\Upsilon\to
\g+a)<10^{-(5-6)}$\,\cite{:2008hs}, which is given by $f_R \gtrsim
10^3$\,GeV.  Furthermore, since we are considering the R-axion with
mass heavier than a few tens of MeV, it is free from the astrophysical
constraints.%
\footnote{As discussed in Ref.\,\cite{Goh:2008xz}, the R-axion can be
  detected at the LHC experiment if the decay constant is in tens of TeV
  range which makes the R-axion mainly decay into a muon pair.}

\section{Conclusion}
In this paper, we have revisited the possibility of the dark matter in
the SUSY breaking sector, in the light of the recent cosmic ray
experiments.  In our model, the dark matter is identified as a
pseudo-Nambu--Goldstone mode in the SUSY breaking sector with a mass in
a TeV range which makes it possible to interpret the observed bump in
the $e^++e^-$ flux at ATIC/PPB-BETS experiments.  Interestingly, the
observed dark matter density requires an existence of a narrow
resonance through which the dark matter annihilates, which results in
a large effective boost factor (in the case of the unphysical pole).
In addition, the dominant final state of the annihilation process is a
pair of the R-axions each of which decays into a pair of light
leptons. Therefore, the Nambu--Goldstone dark matter model is quite
favorable to explain the PAMELA anomaly.

Several comments are in order.  In the model of the Nambu--Goldstone
dark matter, the SUSY breaking scale is around 30\,TeV.  Thus,
the model is accompanied by the gravitino with a mass in a ten eV
range.  The gravitino with such a small mass is attractive, since it
causes no problem in cosmology and astrophysics\,\cite{Moroi:1993mb,Viel:2005qj}.

In the Nambu--Goldstone dark matter scenario, the dark matter mass is
controlled by the degree of the explicit breaking of the approximate
global $SO(6)$ symmetry, {\it i.e.}\/, the difference between $\l$ and $\l'$
(see Eq.\,(\ref{eq:mDM})).  One may attribute the origin of the tuning
between $\l$ and $\l'$ to a conformal dynamics at high energy
scales.  As discussed in Ref.\,\cite{Ibe:2005pj}, the conformal extensions of the vector-like SUSY
breaking model possess an IR-fixed point where the global symmetry
is enhanced.  In such models, even if $|\l' -\l| = O(1)$ at a high
energy scale, the couplings flow to the IR-fixed point in the course
of the renormalization group evolution and end up with $|\l - \l'| \ll1$,
 at the scale of the SUSY breaking.  Thus, if we assume that the
SUSY breaking sector was in a conformal regime at higher energy scales
than the SUSY breaking scale, we can explain the lightness of the dark
matter compared with the scale of the SUSY breaking.

\section*{Acknowledgements}
The work of M.~I. was supported by the U.S. Department of Energy under
contract number DE-AC02-76SF00515.  The research of Y.~N. is supported
in part by NSF grant PHY-0555662 and the UC Berkeley Center for
Theoretical Physics.  The work of H.M. and T.T.Y. was supported in
part by World Premier International Research Center Initiative (WPI
Initiative), MEXT, Japan. The work of H.M. was also supported in part
by the U.S. DOE under Contract DE-AC03-76SF00098, and in part by the
NSF under grant PHY-04-57315.

\appendix
\section{Coleman-Weinberg Potential of IYIT model}\label{sec:cw}
Here, we show the detailed analysis of the Coleman-Weinberg potential
of the flaton $S_{0}$ (see also Ref.\,\cite{Chacko:1998si}.)
The classical flat direction $S_{0}$ is lifted by a one-loop correction
via the interaction $W= \l S_{0} M_{a}^{2}$.  Using the notation
$\sigma = \lambda S_{0}$, $x = \lambda'\, \L_{\rm dyn} $, $y =
\lambda\, \L_{\rm dyn}$, the mass matrix for the fermions is
\begin{equation}
  M_f = \left(
    \begin{array}{cc}
      -\sigma & x\\
      x & 0
    \end{array} \right)
\end{equation}
and for the bosons
\begin{equation}
  M_b^2 = \left(
    \begin{array}{cccc}
      x^2 + \sigma^2 & -x \sigma & - y^2 & 0\\
      -x\sigma & x^2 & 0 & 0\\
      -y^2 & 0 & x^2 + \sigma^2 & - x \sigma\\
      0 & 0 & -x \sigma & x^2
    \end{array} \right)
\end{equation}
The eigenvalues of the fermion mass-squared matrix are
\begin{equation}
  m_f^2 = \frac{1}{2} \left( 2x^2 + \sigma^2 \pm \sigma
    \sqrt{4x^2+\sigma^2} \right),
\end{equation}
while for bosons
\begin{eqnarray}
  m_b^2 &=& \frac{1}{2} \left( 2x^2 + \sigma^2 - y^2 
   \pm \sqrt{4 x^2 \sigma^2 + \sigma^4 - 2 y^2 \sigma^2 + y^4}
  \right), \\
  & & \frac{1}{2} \left( 2x^2 + \sigma^2 + y^2 \pm \sqrt{4 x^2 \sigma^2 + \sigma^4 + 2 y^2 \sigma^2 + y^4} \right).
\end{eqnarray}
Using this spectrum, we can compute the Coleman--Weinberg potential. 
\begin{eqnarray}
    \Delta V_{CW}
  = \frac{5}{64\pi^{2}}{\rm STr}\, m^4 \ln m^2 \ ,
\end{eqnarray}
where a factor $5$ comes from the number of $M_{a}$.
Expanding it up to second order in $\sigma$, we obtain
\begin{eqnarray}
  {\rm STr}\, m^4 \ln m^2
  &=& - 4x^4 \ln x + (x^2-y^2)^2 \ln(x^2-y^2)
  + (x^2+y^2)^2 \log(x^2+y^2) \nonumber \\
  & & 
  + \frac{2}{y^2} \left( (x^2+y^2)^2 \log(x^2+y^2) - (x^2-y^2)^2
    \log(x^2-y^2) \right. \nonumber \\
  & & \left. - 4x^2 y^2 \log(x^2) - 2x^2 y^2\right) \sigma^2
  + O(\sigma^4) .
\end{eqnarray} 
 Since $y<x$ is needed to avoid tachyon, we take the small $y$ limit as
\begin{eqnarray}
  {\rm STr}\, m^4 \ln m^2
  &=& y^4 (3+4\log(x)) + \frac{4y^4}{3x^2} \sigma^2,
\end{eqnarray}
which is a good approximation even as $y \rightarrow x$.  Within this
approximation, the mass term for $S_0$ from the Coleman--Weinberg
potential is
\begin{eqnarray}
  \Delta V_{CW} = \frac{5}{64\pi^2}
  \frac{4 (\lambda\, \L_{\rm dyn})^4}{3(\lambda'\, \L_{\rm dyn})^2} |\lambda S|^2 
= \frac{5}{3(4\pi)^{2}} \frac{\l^{6}}{\l'^{2}}\L_{\rm dyn}^{2} |S_0|^{2}.
\end{eqnarray}.%

Therefore, we obtain the mass of the flat
direction,\footnote{Corresponding K\"ahler potential corrections to
  reproduce $\Delta V_{CW}$ is given by
\begin{equation}
  \Delta K = -\frac{5}{3(4\pi)^2} \frac{\l^4}{4\l'^2\L_{\rm dyn}^2}
  (S_0^\dagger S_0)^2.
\end{equation}
}
\begin{eqnarray}
  m_{S_{0}} \simeq \sqrt{\frac{5}{3}}\frac{\l^{3}}{(4\pi)\,\l'} \L_{\rm dyn}.
\end{eqnarray}
Notice that the flat direction is also lifted by higher dimensional
terms of $S_{0}$ in the K\"ahler potential which is suppressed by the
dynamical scale $4\pi\L_{\rm dyn}$.  However, the flat direction mass is
dominated by the one-loop contribution analyzed here, since the fields
circulating in the loop is much lighter than $4\pi\L_{\rm dyn}$.

\section{R-breaking in gauged IYIT model\label{sec:mason}}
In the main text, we have discussed how R-symmetry breaking in the
hidden sector drastically changes the decay process of SSDM
scenario. In this appendix, we study the R-symmetry breaking of the
IYIT model with additional $U(1)$ gauge symmetry. We embed $SO(2)
\times SO(4)$ in the original $SO(6)$ global symmetry of the IYIT
model, where $SO(2) = U(1)$ is gauged. The dark matter candidate $M_a$
lies in vector representation of $SO(4)$ ($a=1,\cdots 4$).

The low-energy effective superpotential of the gauged IYIT model is
given by%
\footnote{When we set $M_a = S_a=0$, our model is equivalent to
  $k\to\infty$ limit of the model with $W=x S_+ M_- + x S_- M_+ +
  kX(M_+M_- - \L_{\rm dyn}^2)$ which was studied in
  \cite{Dine:2006xt}. }
\begin{eqnarray}
W = xS_+ M_- + x S_- M_+ + y S_a M_a 
\end{eqnarray}
with the constraint $M_+M_- +\frac{1}{2}M_aM_a -\L_{\rm dyn}^2 =
0$. The subscript $\pm$ denotes the $U(1)$ charge of the chiral
superfields.
We parametrize the solution of the deformed moduli constraint as
\begin{eqnarray}
M_+ = \L_{\rm dyn}\, e^{\phi/\sqrt{2}\L_{\rm dyn}} \ , 
\ \ M_- = \L_{\rm dyn}\, e^{-\phi/\sqrt{2}\L_{\rm dyn}} \   \ . 
\end{eqnarray}

In these variables, the leading order K\"ahler potential is
canonically normalized:
\begin{eqnarray}
K = |S_+|^2 + |S_-|^2 + |\phi|^2 + \sum_a (|M_a|^2 + |S_a|^2) + \cdots \ ,
\end{eqnarray}
where non-canonical K\"ahler potential may be neglected when
$F_{S_\pm} = \lambda\L_{\rm dyn}^2\ll \L_{\rm dyn}^2$.  On the other
hand, the superpotential can be written as
\begin{eqnarray}
W &=& mv\left(S_+ e^{-{\phi}/{\sqrt{2}\L_{\rm dyn}}} + S_- e^{{\phi}/{\sqrt{2}\L_{\rm dyn}}}\right)\sqrt{\L_{\rm dyn}^2 - \frac{1}{2}M_aM_a} 
\end{eqnarray}

The tree level vacua have moduli space spanned by $\phi= M_a = S_a=
0$, $S_+=S_- = \sigma$. At the tree level, massless degrees of freedom are
one R-axion, one real modulus $\sigma = (\mathrm{Re} [S_+] +
\mathrm{Re} [S_-])/2$ and one goldstino after gauging away the $U(1)$
Nambu--Goldstone boson at generic points of the moduli space. The
R-axion remains massless in the field theory limit, while the
pseudo-modulus $\sigma$ will acquire a quantum potential, whose shape
and resulting VEV determines whether the R-symmetry is broken.

\begin{figure}[t]
    \begin{center}
    \includegraphics[width=0.4\linewidth]{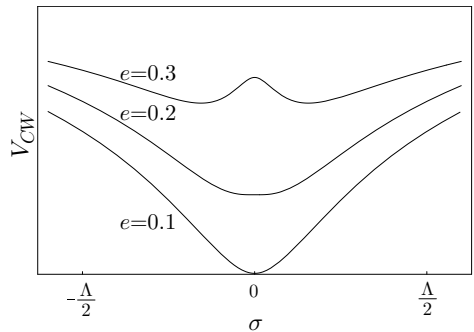}
    \end{center}
    \caption{The Coleman-Weinberg potential for the $U(1)$ gauged IYIT
      model.  In the figure, we have assumed $\l =0.5$, $\l'=1$, and
      the gauge coupling constant as shown in the figure.  }
    \label{fig:CW}
\end{figure}

To see the R-symmetry breaking, we compute the Coleman-Weinberg
potential $V(\sigma) =(1/64\pi^2)\, \mathrm{STr}\,m^4(\sigma) \log
{m^2(\sigma)}$.  In Fig.\,\ref{fig:CW}, we plot the Coleman-Weinberg
potential for a given $U(1)$ gauge coupling constant.  The figure
shows that the symmetry enhancement point $\s=0$ becomes the local
maximum and the Coleman-Weinberg potential develops a minimum at
$\sigma\neq 0$ for a larger value of the additional gauge coupling
constant.  Thus, our $U(1)$ gauged IYIT model serves as a perturbative
model of R-breaking hidden sector with hidden dark matter. The
R-breaking depends on the $U(1)$ coupling constant.
\section{F-flaton Decay into the SSM particle}\label{sec:ssm}
In this appendix, we consider the decay modes of the flaton into the
SSM particles.  Since we are assuming the model with gauge mediation,
the flaton couples to the SSM fields as the results of the mediation
effects.  For example, the effective coupling between the flaton and
the gauginos is given by a Yukawa interaction;
\begin{eqnarray}
\label{eq:gaugino}
 {\cal L}_{\rm eff} \simeq \frac{1}{2}\, \frac{m_i }{f_{R}}
 \left(1+O\left(\frac{f_{R}^{2}}{F_{S_0}}\right) \right)
 \, 
s \,\l^{i}\l^{i} 
 + h.c.,
\end{eqnarray}
where $m_{i}$ denotes the gaugino mass and $i$ runs the SSM gauge
groups.  Notice that the leading term in the above effective coupling
is model independent as long as the messenger sector possesses the
R-symmetry.  On the other hand, the coupling between the flaton and
the sfermions depends on the messenger sector even if it is
R-symmetric, and is given by,
\begin{eqnarray}
\label{eq:scalar}
{\cal L}_{\rm eff} = \left.\frac{\partial m_{\tilde f}^{2}}{\partial s}\right|_{s=0} \times \,s\,\tilde f \tilde f,
\end{eqnarray}
where the model dependent coefficient ${\partial m_{\tilde
    f}^{2}}/{\partial s}$ satisfies 
\begin{eqnarray}
 \left.\frac{\partial m_{\tilde f}^{2}}{\partial s}\right|_{s=0} \leq \,\frac{m_{\tilde f}^{2}}{f_{R}}.
\end{eqnarray}
From these interactions, the flaton decays into a pair of the SSM
particles.  For instance, the decay rate into a pair of the gluinos
are given by
\begin{eqnarray}
\label{eq:gaugino2}
 \G_{s\to \tilde g\tilde g} \simeq \frac{1}{4\pi} \left( \frac{m_{\tilde g}}{m_{s}} \right)^{2}\frac{m_{s}^{3} }{f_{R}^{2}}.
\end{eqnarray}
Therefore, depending on the spectrum of the SSM and the dark matter
($m_{\rm DM}\simeq m_{s}/2$), the branching ratio of the flaton into
the SSM particles can be suppressed.  Notice that the branching ratio
into the gravitino pair is highly suppressed\,\cite{Ibe:2006rc}.

\end{document}